\documentclass[sigconf,nonacm,balance=false]{acmart}
\AtBeginDocument{%
  }

\setcopyright{none}
\acmYear{2027}
\acmConference[KDD '27]{Proceedings of the 33rd ACM SIGKDD Conference on Knowledge Discovery and Data Mining V.1}{August 01--05, 2027}{San Jose, CA, USA}
\acmBooktitle{Proceedings of the 33rd ACM SIGKDD Conference on Knowledge Discovery and Data Mining V.1 (KDD '27), August 01--05, 2027, San Jose, CA, USA}
\acmDOI{10.1145/3862868.3871559}
\acmISBN{979-8-4007-3134-1/2027/08}

\usepackage{makecell}
\usepackage{pifont}
\usepackage{xcolor}
\usepackage{bbding}
\definecolor{DarkGreen}{RGB}{0, 128, 0}  
\usepackage[table]{xcolor}
\usepackage{geometry}
\usepackage{booktabs}
\usepackage{tabularx}
\usepackage{multirow}
\usepackage{subcaption}
\definecolor{lightgreen}{HTML}{E8F5E9}
\definecolor{lightblue}{HTML}{E1F5FE}
\usepackage[most]{tcolorbox}
\newtcolorbox{listingbox}[1][]{
  enhanced,
  breakable,
  colback=lightblue,
  colframe=black!80,
  boxrule=0.4pt,
  arc=1pt,
  left=4pt,
  right=4pt,
  top=4pt,
  bottom=4pt,
  fontupper=\small,
  fonttitle=\bfseries,
  #1
}
\begin{document}
\title{SciEGQA: A Dataset for Scientific Evidence-Grounded Question Answering and Reasoning}

\author{Wenhan Yu}
\email{yuwenhan@buaa.edu.cn}
\affiliation{%
  \institution{Beihang University}
  \city{Beijing}
  \country{China}
}

\author{Zhaoxi Zhang}
\email{zhaoxizhang25@stu.pku.edu.cn}
\affiliation{%
  \institution{Peking University}
  \city{Beijing}
  \country{China}
}

\author{Wang Warren Chen}
\email{wchen22@connect.hku.hk}
\affiliation{%
  \institution{The University of Hong Kong}
  \city{Hong Kong}
  \country{China}
}

\author{Guanqiang Qi}
\email{qiguanqiang@gmail.com}
\affiliation{%
  \institution{Baidu Inc.}
  \city{Beijing}
  \country{China}
}

\author{Weikang Li}
\email{wavejkd@pku.edu.cn}
\authornote{Project leader and corresponding author}
\affiliation{%
  \institution{Peking University}
  \city{Beijing}
  \country{China}
}
\affiliation{%
  \institution{Baidu Inc.}
  \city{Beijing}
  \country{China}
}

\author{Lei Sha}
\authornotemark[1]
\email{shalei@buaa.edu.cn}
\affiliation{%
  \institution{Beihang University}
  \city{Beijing}
  \country{China}
}

\author{Deguo Xia}
\email{xiadeguo@baidu.com}
\affiliation{%
  \institution{Tsinghua University}
  \city{Beijing}
  \country{China}
}
\affiliation{%
  \institution{Baidu Inc.}
  \city{Beijing}
  \country{China}
}

\author{Jizhou Huang}
\email{huangjizhou01@baidu.com}
\affiliation{%
  \institution{Baidu Inc.}
  \city{Beijing}
  \country{China}
}

\renewcommand{\shortauthors}{Wenhan Yu et al.}

\begin{abstract}
Scientific documents contain complex multimodal structures, which makes evidence localization and scientific reasoning in Document Visual Question Answering particularly challenging. However, most existing benchmarks evaluate models only at the page level without explicitly annotating the evidence regions that support the answer, which limits both interpretability and the reliability of evaluation.
To address this limitation, we introduce SciEGQA, a scientific document question answering and reasoning dataset with semantic evidence grounding, where supporting evidence is represented as semantically coherent document regions annotated with bounding boxes. SciEGQA consists of two components: a human-annotated fine-grained benchmark containing 1,623 high-quality question--answer pairs, and a large-scale automatically constructed training set with over 30K QA pairs generated through an automated data construction pipeline.
Extensive experiments on a wide range of Vision-Language Models (VLMs) show that existing models still struggle with evidence localization and evidence-based question answering in scientific documents. Training on the proposed dataset significantly improves the scientific reasoning capabilities of VLMs. The project page is available at \url{https://yuwenhan07.github.io/SciEGQA-project/}.

\end{abstract}




\maketitle

\section{Introduction}
With the rapid development of Multimodal Large Language Models (MLLMs)~\cite{li2024improving,wu2024next,yin2024survey,caffagni2024revolution}, Vision-Language Models (VLMs) have achieved remarkable progress in diverse tasks such as image understanding, visual question answering, and multimodal reasoning~\cite{wang2024multimodal}.
However, compared with natural images, scientific documents present fundamentally different challenges. 
A page of a scientific paper often contains dense multimodal content, including textual paragraphs, tables, figures, and cross-page references~\cite{wang2024charxiv,li2024multimodal}. 
These characteristics collectively make Document Visual Question Answering (DocVQA)~\cite{mathew2021docvqa} particularly challenging, as models must not only accurately localize the relevant evidence regions but also integrate information across modalities to perform reasoning. In this work, we study scientific documents directly as rendered page images rather than as OCR transcripts or parsed text. This image-based setting preserves both visual content and spatial layout~\cite{dasigi2021dataset,lee2023qasa,singh2024scidqa,wang2025mesaqa}.

In recent years, a variety of DocVQA benchmarks have been proposed to evaluate document understanding capabilities~\cite{mathew2021docvqa,tito2023hierarchical,mathew2022infographicvqa,saikh2022scienceqa}. Nevertheless, existing scientific document QA datasets still exhibit several limitations in how supporting evidence is represented.
Datasets that only provide \textbf{doc/page-level} evidence annotations are often too coarse-grained for scientific documents, where a single page may contain multiple dense technical elements such as paragraphs, figures, and tables. As a result, page-level supervision cannot precisely identify the specific regions that support an answer~\cite{ma2024mmlongbench,li2024multimodal,singh2024scidqa,baumgartner2025peerqa}.
Furthermore, some datasets construct question answering tasks around \textbf{specific structured components}, such as a single table or figure~\cite{wang2024charxiv,masry2022chartqa,pramanick2024spiqa,yue2025anafig}. While useful for studying individual modalities, these settings isolate individual visual elements from the broader document context and therefore cannot capture the complete semantic information of the document.
In contrast, \textbf{token-level} annotations, although more fine-grained, often fragment semantically coherent units and lose important contextual information, making systematic reasoning difficult and reducing interpretability~\cite{giovannini2025boundingdocs,tanaka2021visualmrc,van2023document}. 

However, incorporating precise evidence region localization into the reasoning process has been shown to be effective in visual question answering~\cite{zheng2025deepeyes,hong2025deepeyesv2,zhang2025thyme}, yet this capability is rarely systematically evaluated in existing scientific DocVQA benchmarks. 
Therefore, an effective evidence representation should support \textbf{semantic region-level grounding}, which preserves semantic completeness while enabling precise spatial localization of the supporting evidence.

To address these limitations, we introduce \textbf{SciEGQA}, a visual question answering dataset for scientific documents that incorporates \textit{semantic evidence grounding}. In SciEGQA, the supporting evidence for each answer is represented as \textbf{semantically coherent document regions} annotated with bounding boxes. This design provides an intermediate granularity between page-level and token-level annotations, enabling precise spatial localization while preserving semantic completeness, which is more suitable for multimodal reasoning over scientific documents.

SciEGQA consists of two complementary components: a \textbf{human-annotated benchmark} constructed from 80 scientific papers across eight research domains with 1,623 high-quality QA pairs, and a \textbf{large-scale automatically constructed training set} generated from more than 3.6K papers containing over 30K QA pairs. The dataset supports three reasoning scenarios with increasing complexity: \textbf{Single-Page Single-Region (SPSR)}, \textbf{Single-Page Multi-Region (SPMR)}, and \textbf{Multi-Page Multi-Region (MPMR)}.
We further demonstrate through supervised fine-tuning experiments that the automatically constructed training set provides effective supervision for improving scientific reasoning capability.

Based on SciEGQA, we conduct comprehensive evaluations on a wide range of state-of-the-art vision-language models. Experimental results reveal that even the most advanced models still struggle with evidence localization and evidence-based reasoning in scientific documents. In particular, most models achieve less than $40\%$ IoU in evidence grounding tasks, highlighting the challenging nature of scientific document understanding.







The main contributions of this work are summarized as follows:
(1) We introduce \textbf{SciEGQA}, a fine-grained benchmark and training set for evidence-grounded question answering and reasoning on scientific documents.
(2) We propose \textbf{semantic region-level evidence grounding}, an intermediate annotation granularity that preserves semantic completeness while enabling precise spatial localization for reasoning.
(3) We propose a \textbf{Grounding-Crop-then-Answer} evaluation protocol that disentangles evidence localization from reasoning ability.

\begin{table}[t]
\centering
\caption{Comparison of Scientific DocVQA benchmarks across doc/page-level, component-level, and token-level datasets and our SciEGQA benchmark.}
\label{tab:benchmark-comparison}
\setlength{\tabcolsep}{4pt}
\resizebox{\linewidth}{!}{
\begin{tabular}{l c c c c c}
\toprule
\textbf{Dataset} & \textbf{\#QA} & \textbf{Evidence} & \textbf{Multi-page} & \textbf{BBox} & \textbf{\makecell{Evidence \\ Granularity}} \\

\midrule
DocVQA~\cite{mathew2021docvqa} & 5.19k & \textcolor{DarkGreen}{\Checkmark} & \textcolor{red}{\XSolidBrush}  & \textcolor{red}{\XSolidBrush}  & Page Level \\
MP-DocVQA~\cite{tito2023hierarchical} & 5.02k & \textcolor{DarkGreen}{\Checkmark} & \textcolor{DarkGreen}{\Checkmark} & \textcolor{red}{\XSolidBrush}  & Page Level \\
TAT-DQA~\cite{zhu2022towards} & 1.66k & \textcolor{DarkGreen}{\Checkmark} & \textcolor{red}{\XSolidBrush}  & \textcolor{red}{\XSolidBrush}  & Page Level \\
SciDQA~\cite{singh2024scidqa} & 2.94k & \textcolor{red}{\XSolidBrush} & \textcolor{DarkGreen}{\Checkmark}  & \textcolor{red}{\XSolidBrush}  & Doc Level \\
PeerQA~\cite{baumgartner2025peerqa} & 0.57k & \textcolor{red}{\XSolidBrush} & \textcolor{DarkGreen}{\Checkmark}  & \textcolor{red}{\XSolidBrush}  & Doc Level \\
MMLongBench-Doc~\cite{ma2024mmlongbench} & 1.09k & \textcolor{DarkGreen}{\Checkmark} & \textcolor{DarkGreen}{\Checkmark} & \textcolor{red}{\XSolidBrush}  & Page Level \\

\midrule
CharXiv~\cite{wang2024charxiv} & 1.32k & \textcolor{DarkGreen}{\Checkmark} & \textcolor{red}{\XSolidBrush}  & \textcolor{red}{\XSolidBrush}  & \small Component (Chart) \\
SPIQA~\cite{pramanick2024spiqa} & 0.66k & \textcolor{DarkGreen}{\Checkmark} & \textcolor{red}{\XSolidBrush}  & \textcolor{red}{\XSolidBrush}  & \small Component (Table) \\
AnaFig~\cite{yue2025anafig} & 10k & \textcolor{red}{\XSolidBrush} & \textcolor{red}{\XSolidBrush}  & \textcolor{red}{\XSolidBrush}  & \small Component (Figure) \\

\midrule
VisualMRC~\cite{tanaka2021visualmrc} & 6.71k & \textcolor{red}{\XSolidBrush}  & \textcolor{red}{\XSolidBrush}  & \textcolor{DarkGreen}{\Checkmark} &  Token Level \\
DUDE~\cite{van2023document} & 41k & \textcolor{red}{\XSolidBrush}  & \textcolor{DarkGreen}{\Checkmark} & \textcolor{red}{\XSolidBrush}  & Token Level \\
BoundingDocs~\cite{giovannini2025boundingdocs} & 4.83k & \textcolor{red}{\XSolidBrush}  & \textcolor{red}{\XSolidBrush}  & \textcolor{DarkGreen}{\Checkmark} & Token Level \\
\midrule
\textbf{SciEGQA (Ours)} & \textbf{1.62k} & \textcolor{DarkGreen}{\Checkmark} & \textcolor{DarkGreen}{\Checkmark} & \textcolor{DarkGreen}{\Checkmark} & \textbf{Semantic Region Level} \\
\bottomrule
\end{tabular}
}
\end{table}
\section{Related Work}
\label{sec:related_work}

\begin{figure*}
    \centering
    \includegraphics[width=\linewidth]{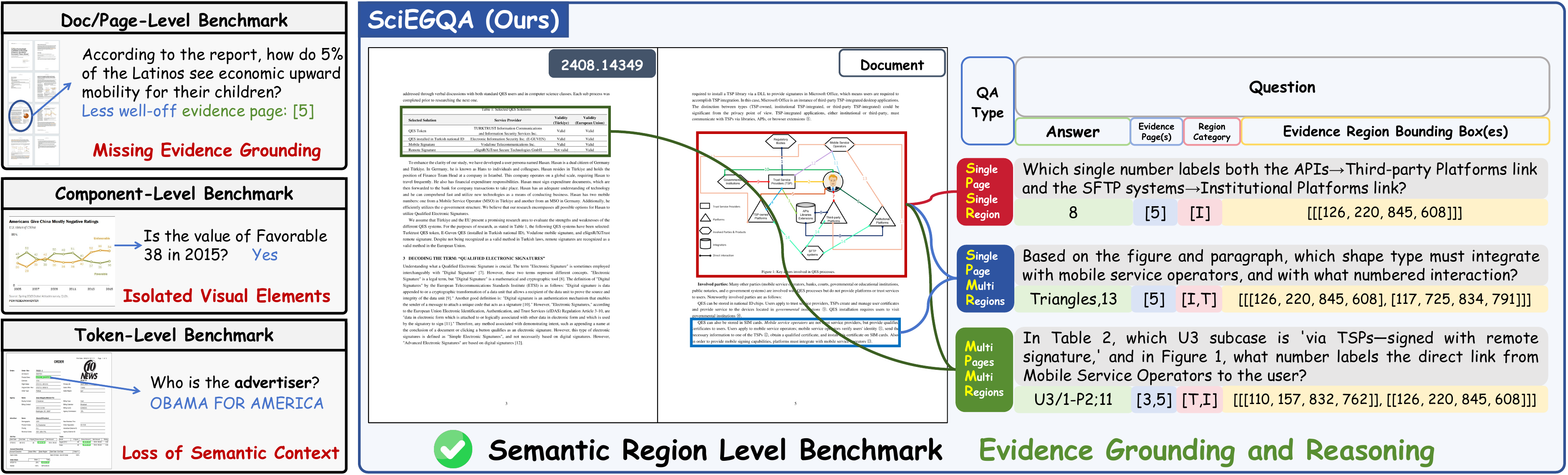}
    \caption{
    Comparison of evidence annotation granularity in document QA datasets. Existing benchmarks provide page-level QA, token-level grounding, or component-level reasoning. In contrast, SciEGQA supports page-level question answering with semantic region grounding via bounding-box evidence.
    }
    \Description{Three examples on the left show evidence marked at page, component, or token level. On the right, boxes link a diagram, a paragraph, and a table on two pages to questions requiring one region, multiple regions on one page, or regions across pages.}
    \label{fig:relate_work}
\end{figure*}

Document Visual Question Answering (DocVQA) has attracted significant attention in recent years, leading to the development of various benchmark datasets for evaluating document understanding capabilities. While many datasets focus on general document types such as receipts, forms, and infographics, scientific documents introduce additional challenges due to their domain-specific characteristics and complex multimodal structures. As illustrated in Figure~\ref{fig:relate_work} and Table~\ref{tab:benchmark-comparison}, existing Scientific DocVQA datasets can be broadly categorized into three types: doc/page-level datasets, structured component-level datasets, and token-level datasets.

\paragraph{Doc/Page-level datasets.}
Early datasets such as DocVQA~\cite{mathew2021docvqa} primarily focus on question answering over single-page documents. MP-DocVQA~\cite{tito2023hierarchical} extends this setting to multi-page scenarios, introducing more complex challenges such as cross-page reasoning and long-context understanding. In addition, datasets including SciDQA~\cite{singh2024scidqa}, PeerQA~\cite{baumgartner2025peerqa} and TAT-DQA~\cite{zhu2022towards} further explore question answering over scientific paper-style documents. More recently, benchmarks such as MMLongBench-Doc~\cite{ma2024mmlongbench} have been proposed to systematically evaluate long-document understanding capabilities of multimodal models. However, these datasets typically provide only page-level evidence annotations, lacking precise localization of fine-grained answer-supporting regions.

\paragraph{Structured component-level datasets.}
Another line of work further focuses on structured visual components with rich semantic information in documents, such as charts, tables, and figures. Representative benchmark datasets include SPIQA~\cite{pramanick2024spiqa}, CharXiv~\cite{wang2024charxiv}, and AnaFig~\cite{yue2025anafig}. These datasets typically construct question answering tasks around specific document components, particularly figures or charts extracted from scientific papers or technical reports. While this setting facilitates fine-grained reasoning over individual visual components, it often isolates these elements from the broader document context, making it difficult to evaluate comprehensive document understanding and cross-component information integration.

\paragraph{Token-level datasets.}
Another line of work adopts token-level supervision by aligning answers with OCR-extracted text tokens. Datasets such as VisualMRC~\cite{tanaka2021visualmrc}, DUDE~\cite{van2023document}, and BoundingDocs~\cite{giovannini2025boundingdocs} support fine-grained evidence localization at the textual token level. Although this approach provides precise spatial localization signals, the annotations are typically limited to individual tokens or short phrases, lacking complete semantic context and making it difficult to capture hierarchical structures and compositional semantics within documents.

Overall, existing datasets still exhibit notable limitations in fine-grained document modeling. Doc/Page-level methods lack precise localization, token-level methods lack semantic completeness, and component-level methods often focus on isolated visual elements rather than holistic document reasoning. These limitations are particularly evident in scientific documents, where answering questions often requires integrating information across text, figures, tables, and multiple document regions. 
To address these challenges, we introduce the \textbf{SciEGQA} dataset, which models supporting evidence as \textbf{semantic regions} annotated with bounding boxes, enabling both precise evidence localization and multimodal reasoning across document regions.

\section{SciEGQA Dataset}
\begin{figure}
    \centering
    \includegraphics[width=0.9\linewidth]{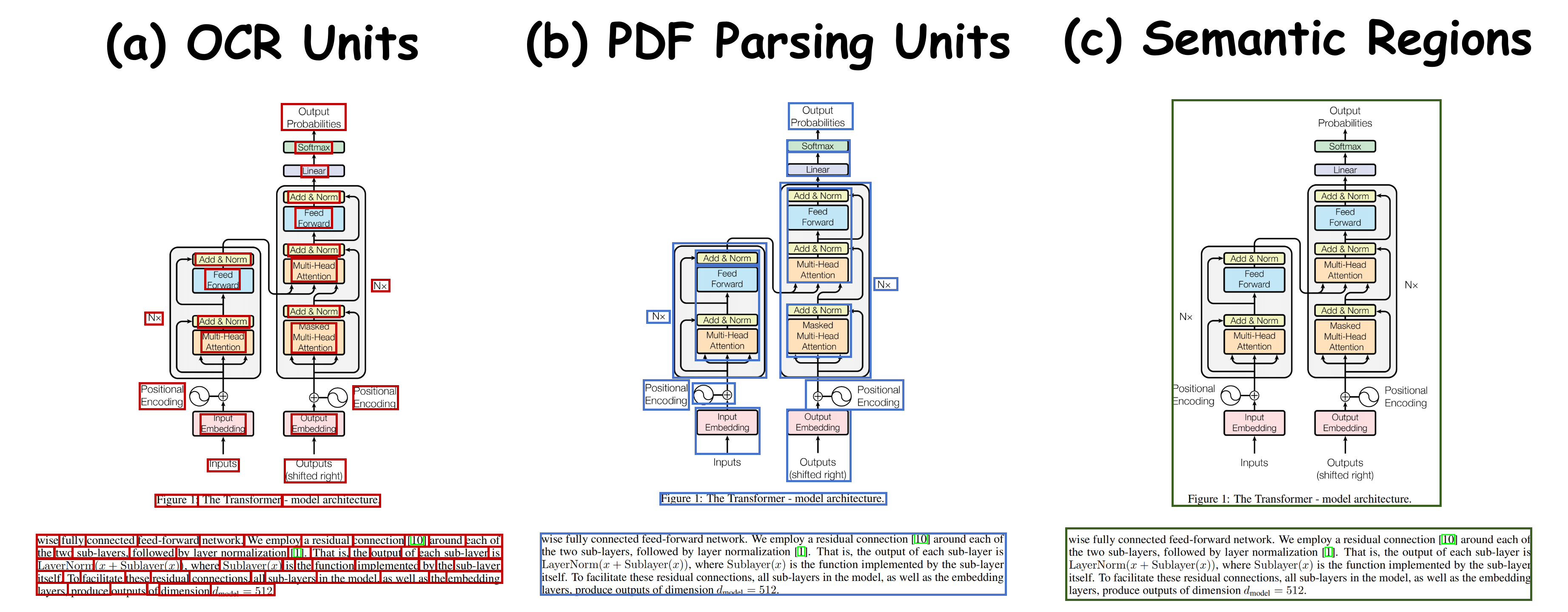}
    \caption{Comparison of OCR, PDF parsing, and semantic-region representations. OCR produces fine-grained text boxes, PDF parsing organizes content according to file objects, while semantic regions preserve coherent evidence units.}
    \Description{The same Transformer diagram and paragraph appear in three panels. OCR boxes surround individual words and labels, PDF parsing boxes surround separate objects, and semantic-region boxes enclose the diagram with its caption and the paragraph as two complete units.}
    \label{fig:method-difference}
\end{figure}

\subsection{Semantic Regions and Evidence Units}

We define a \textbf{Semantic Region (SR)} as the smallest spatially contiguous document region that preserves a complete semantic unit. An SR is a reusable document-level grounding unit rather than a geometry-driven layout block or a question-specific crop.

As shown in Figure~\ref{fig:method-difference}, OCR and PDF-parsing outputs do not reliably form complete semantic units~\cite{smith2007overview,blecher2308nougat}. OCR omits non-textual cues, while object-level parsing can lose relationships encoded by page layout~\cite{clark2016pdffigures,luo2024layoutllm,huang2022layoutlmv3}. SR boundaries are instead determined by semantic completeness.

SR determination follows two principles: \textit{semantic completeness} and \textit{minimality}. A region must contain all content needed to interpret the unit while excluding adjacent content that is not required.

Text SRs are paragraphs or lists. Figure SRs include necessary captions; table SRs include the title, body, and notes; and equations include the surrounding text needed for interpretation.

Unrelated columns are not merged. Cross-page evidence is represented by multiple page-local SRs, each preserving a complete semantic unit. These rules preserve necessary context while maintaining precise localization.

For a question-answer pair, an \textbf{Evidence Unit (EU)} is an SR selected because it provides information needed to derive the answer. Thus, SR denotes a reusable document unit, whereas EU denotes its question-specific evidential role.

A question may require one or multiple EUs. Its supporting evidence is represented as a set of EUs, including multiple boxes when reasoning spans regions or pages.

\subsection{Overview of the SciEGQA Dataset}

SciEGQA is designed to evaluate models on the task of \textbf{scientific document understanding with evidence grounding}. 
Unlike conventional DocVQA datasets that focus only on predicting the correct answer~\cite{ma2024mmlongbench,mathew2021docvqa,masry2022chartqa}, SciEGQA explicitly models the supporting evidence required for answering each question.

Formally, each sample $S$ in SciEGQA can be represented as
\[
S = (D, Q, A, T, E),
\]
where $D$ is the source document, $Q$ is the question, $A$ is the answer, and $T$ is the reasoning type. Let $R(D)$ denote the collection of annotated SRs in $D$.

The evidence set satisfies $E=\{e_i\}_{i=1}^{n}\subseteq R(D)$, where each EU $e_i=(p_i,b_i,c_i)$ records its page $p_i$, bounding box $b_i=(x_1,y_1,x_2,y_2)$, and semantic category $c_i$ (e.g., text, figure, or table).

SciEGQA is designed with three key principles:
\textbf{Semantic completeness.}
Each EU inherits the complete and minimal boundary of its underlying SR rather than using fragmented tokens or a question-specific crop.
\textbf{Multimodal coverage.}
Scientific documents contain diverse modalities including text, tables, and figures. SciEGQA explicitly models evidence across these modalities to support realistic multimodal reasoning scenarios.
\textbf{Reasoning diversity.}
To reflect the complexity of scientific document understanding, SciEGQA includes three reasoning scenarios with increasing difficulty: \textbf{Single-Page Single-Region (SPSR)}, \textbf{Single-Page Multi-Region (SPMR)}, and \textbf{Multi-Page Multi-Region (MPMR)}.
Each question uses one or multiple EUs, depending on whether its answer requires a single SR, multiple SRs on one page, or SRs across pages.

The dataset consists of two complementary components: a \textbf{fine-grained benchmark} designed for rigorous evaluation and a \textbf{large-scale automatically constructed training set} that supports scalable model training.

\subsection{Data Collection}
\label{sec:data_collection}

\begin{figure}
    \centering
    \includegraphics[width=\linewidth]{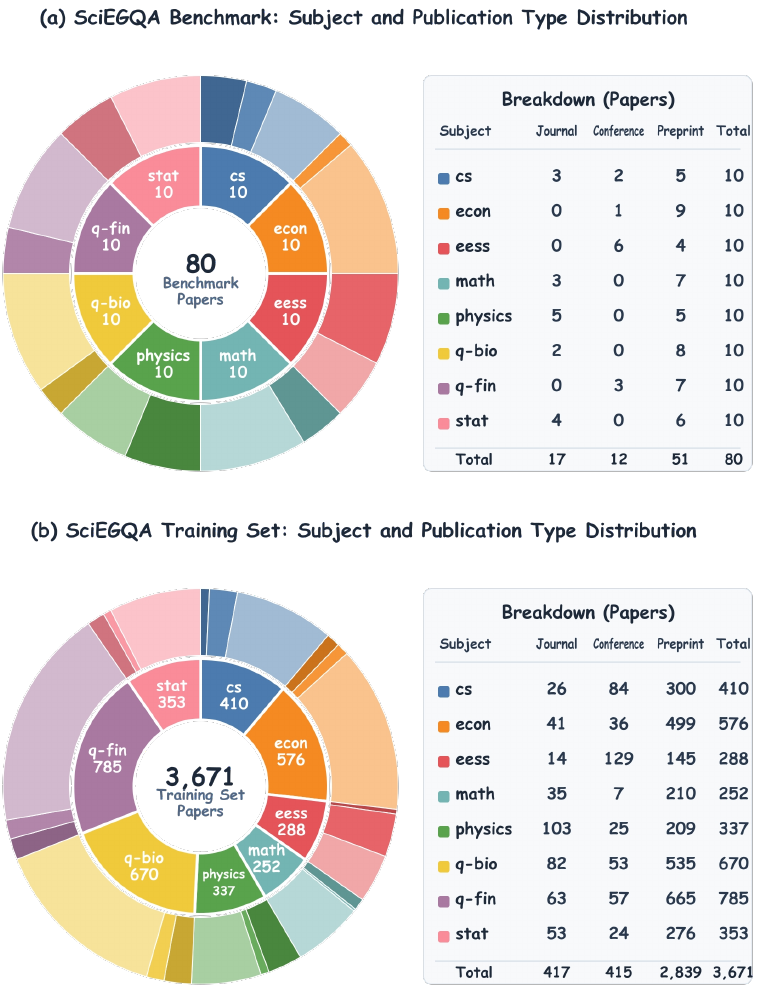}
    \caption{Subject and publication-type distributions of the source papers. (a) The human-annotated benchmark contains 80 papers, balanced with 10 papers per domain. (b) The automatically constructed training set contains 3,671 papers with broader domain and publication-type coverage.}
    \Description{Two nested doughnut charts show research domains in the inner rings and publication types in the outer rings. The benchmark has ten papers per domain, the training set varies by domain, and preprints form the largest publication group in both.}
    \label{fig:document-distributions}
\end{figure}
\begin{table}[t]
\centering
\caption{Dataset scale and reasoning-type distributions of the SciEGQA benchmark and training set.}
\label{tab:dataset_stats}
\resizebox{\linewidth}{!}{
\begin{tabular}{l r r r | r r r}
\toprule
\textbf{Dataset} & \textbf{\#Paper} & \textbf{\#Page} & \textbf{\#QA} & \textbf{SPSR} & \textbf{SPMR} & \textbf{MPMR} \\
\midrule
Benchmark    & 80    & 1,941  & 1,623  & 46.15\% & 34.26\% & 19.59\% \\
Training Set & 3,671 & 42,380 & 30,780 & 37.91\% & 24.41\% & 37.69\% \\
\bottomrule
\end{tabular}
}
\end{table}

\begin{figure*}
    \centering
    \includegraphics[width=0.9\linewidth]{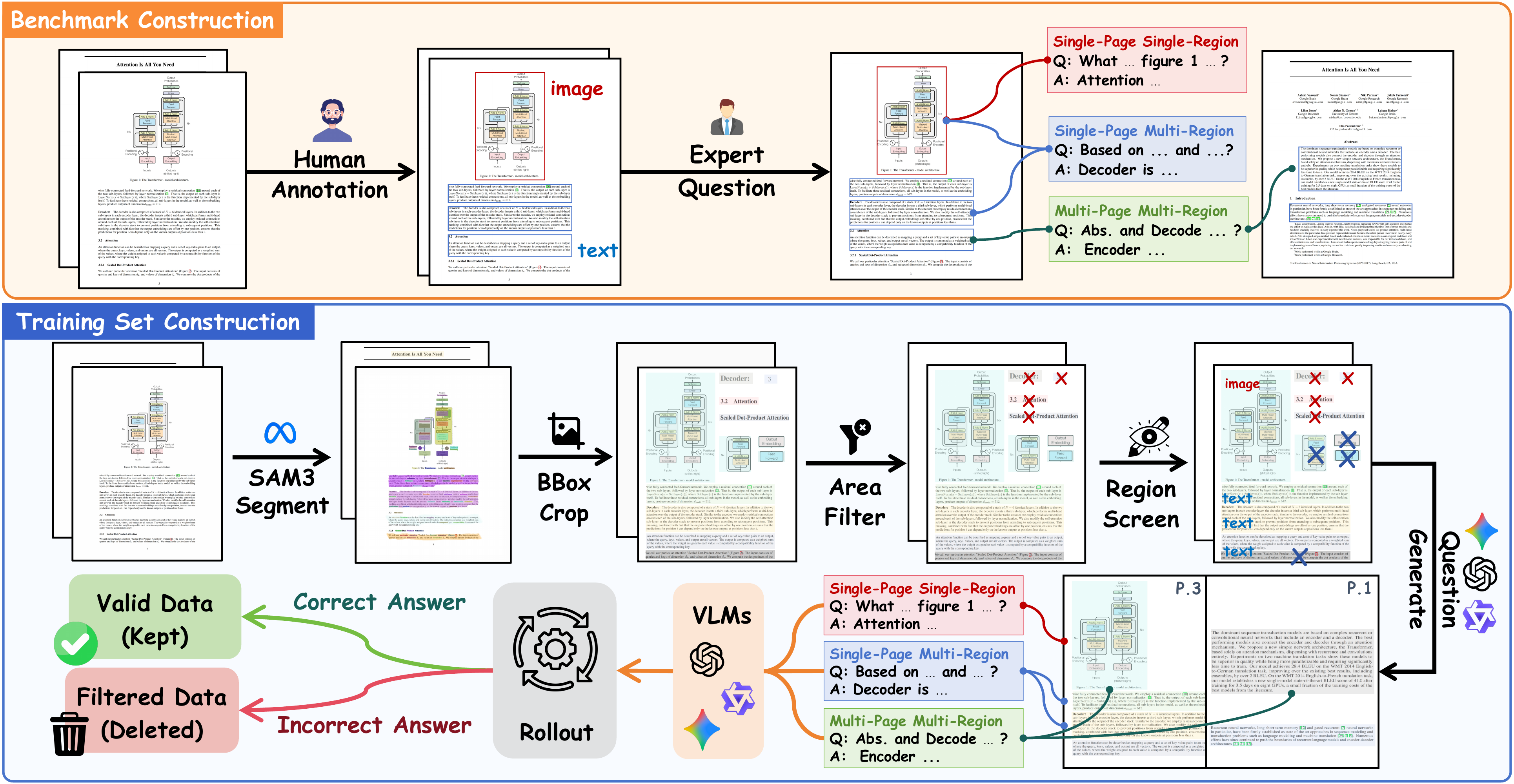}
    \caption{Overview of SciEGQA dataset construction.
The dataset is built through two complementary processes: (top) benchmark construction with human annotation and (bottom) large-scale training set construction via an automated pipeline. }
    \Description{In the top workflow, annotators draw region boxes and experts write questions tied to those regions. In the bottom workflow, SAM3 segments pages, candidates are cropped and filtered, and vision-language models generate and verify question-answer pairs before they are kept or rejected.}
    \label{fig:data-construction-pipeline}
\end{figure*}

SciEGQA collects scientific papers from three publication types: \textbf{journals}, \textbf{conference proceedings}, and \textbf{preprint repositories}. This multi-source composition extends the dataset beyond a single publication venue, ensuring broad coverage across different publication stages and document types.

As illustrated in Figure~\ref{fig:document-distributions} and Table~\ref{tab:dataset_stats}, the dataset spans eight research domains: computer science, economics, electrical engineering and systems science, mathematics, physics, quantitative biology, quantitative finance, and statistics.

For the benchmark, domain experts manually selected \textbf{10 representative papers from each domain}. The selection covers diverse document layouts and publication types, including single-column and double-column formats, as well as journal articles, conference papers, and preprints, to comprehensively represent the common organizational structures and formatting styles of scientific literature. This process resulted in a balanced benchmark consisting of \textbf{80 papers}, with equal representation across all eight domains.

For the automatic construction of the training set, we collected \textbf{3,671 papers} from the same research domains and publication types. Unlike the benchmark, the training corpus preserves the natural distribution of the source data to further improve its scale, diversity, and representativeness.

Figure~\ref{fig:document-distributions} summarizes both the domain distribution and publication-type distribution of the dataset. Specifically, the benchmark contains \textbf{17 journal articles, 12 conference papers, and 51 preprints}, while the training corpus consists of \textbf{417 journal articles, 415 conference papers, and 2,839 preprints}.

During preprocessing, all PDF pages are rendered as PNG images at  \textbf{300 dpi}. This resolution preserves fine-grained visual details, including small text, mathematical symbols, and thin table or figure structures, while keeping the computational and storage costs of region segmentation manageable. We apply the same resolution to all documents to ensure consistent rasterization throughout the annotation and construction pipelines.

\subsection{Fine-grained Benchmark Construction}
As illustrated in the upper part of Figure~\ref{fig:data-construction-pipeline}, we construct a fine-grained benchmark by manually annotating the \textbf{80 sampled scientific papers} described in Section~\ref{sec:data_collection}.

During the annotation process, \textbf{12 annotators} with backgrounds in document understanding or related fields participate in the data construction. 
In addition, \textbf{three domain experts for each research field} are involved in question design and annotation verification, resulting in a total of \textbf{24 domain experts}.

\textbf{Question-independent SR annotation.}
Before any questions are written, annotators inspect each page and draw bounding boxes around candidate SRs. This order prevents region boundaries from being tailored to a particular QA instance.

Annotators apply the completeness and minimality criteria defined above. A box is expanded when a caption, title, note, or explanatory text is required, and is tightened or split when it includes unrelated content.

Cross-page content is never enclosed by one box. Each complete unit on a page is annotated as a separate SR, allowing later questions to combine multiple SRs without sacrificing localization precision.

Furthermore, to preserve semantic completeness, we intentionally exclude fragments that cannot independently convey complete semantics, such as cross-page text fragments and table content whose interpretation depends on missing context. While this design inevitably discards a small amount of cross-page information, it substantially reduces evidence ambiguity, improves annotation consistency, and leads to a more reliable benchmark for evidence grounding evaluation.

\textbf{Boundary agreement and adjudication.}
Each paper is independently annotated by two annotators. Matched SR boxes are compared using Intersection over Union (IoU) to measure boundary agreement.

If the IoU is below \textbf{0.9}, a third annotator reviews both the boundary and semantic completeness of the region. The adjudicated boxes form the finalized SR pool for the document.

\textbf{Evidence-unit selection and QA authoring.}
After the SR pool is finalized, domain experts design questions and select the SRs needed to answer each one. The selected SRs become the EUs for that QA instance.

The questions cover three reasoning scenarios: \textbf{SPSR}, \textbf{SPMR}, and \textbf{MPMR}, with an approximate ratio of \textbf{5:3:2}. 
Each question must be answerable \textit{solely and completely} from its EUs, ensuring a clear correspondence between the answer and its supporting SRs.

This process separates reusable SR annotation from question-specific EU selection. The resulting benchmark supports evaluation of both \textit{document question answering} and \textit{evidence region localization}.

\begin{figure*}
    \centering
    \includegraphics[width=\linewidth]{fig/analysis.pdf}
    \caption{Dataset statistics.
Comparison of the fine-grained benchmark (left) and the automatically constructed training set (right). The figure shows distributions across category–modality combinations, spatial locations of evidence bounding boxes, modality combinations, and region sizes, illustrating the diversity of document types, evidence modalities, and spatial layouts in the dataset.}
    \Description{Side-by-side panels compare the benchmark and training set using page-position heatmaps, modality counts by domain, and bar charts of modality combinations and region sizes. Both heatmaps peak near the horizontal center of the page, and text-only evidence is the most frequent combination in both sets.}
    \label{fig:data-analysis}
\end{figure*}

\subsection{Automatic Training Set Construction}

To construct a large-scale training set, we design an automated data generation pipeline, as illustrated in the lower part of Figure~\ref{fig:data-construction-pipeline}. 
Unlike the benchmark dataset, which is entirely constructed through human annotation, the training set is generated automatically to substantially increase the dataset scale while maintaining reasonable data quality.

\textbf{Region Segmentation.}
For each document page, we use \textbf{SAM3}~\cite{carion2025sam} with its default segmentation settings to generate candidate visual segments directly from the rendered page image. SAM3 serves as a category-agnostic proposal generator rather than a document parser.

Because it operates on visual content, SAM3 preserves both non-textual information and layout cues without imposing predefined document categories. These cues provide a stronger basis for assessing whether candidate segments form complete semantic units.

The resulting segments are not treated as SRs directly. They are proposals to be evaluated and, when necessary, combined in the next stage under the same completeness and minimality criteria used for human annotation.

\textbf{Semantic Region Formation.}
We first filter the segmentation proposals according to their relative page area, removing candidates that occupy less than \textbf{5\%} or more than \textbf{70\%} of the page. This step eliminates degenerate regions that are excessively small, as well as overly large segments that are unlikely to represent valid semantic units. These thresholds are selected through small-scale manual validation to achieve a reasonable balance between retaining meaningful regions and removing anomalous proposals. We then apply semantic filtering with random strong vision–language models and retain proposals that already constitute complete and minimal semantic units.

Spatially adjacent proposals are combined when they are semantically dependent, such as a figure and its caption or a table title and body. Proposals that remain incomplete or noisy are discarded, and unrelated columns are never grouped.

The output of this stage is an automatically constructed SR pool. Thus, SAM3 supplies visual boundaries, while semantic filtering and grouping determine which boundaries satisfy the SR definition.

\textbf{Evidence-set Construction.}
After obtaining the SR pool, the automated pipeline samples one or more semantic regions according to their spatial distribution and semantic relationships to construct candidate evidence sets, covering three reasoning scenarios: single-region, single-page multi-region, and multi-page multi-region reasoning. Once assigned to a QA instance, the selected SRs become the corresponding Evidence Units (EUs), forming \textbf{SPSR}, \textbf{SPMR}, or \textbf{MPMR} evidence sets.

\textbf{Question-Answer Generation.}
Given the selected EUs, we employ both open-source and proprietary vision-language models with strong visual reasoning capabilities, including \textbf{Qwen3-VL}, \textbf{GPT-5}, and \textbf{Gemini 3}, to generate question-answer pairs. During generation, the models are explicitly instructed to construct questions and answers solely based on the provided EUs, ensuring that each answer is grounded in the corresponding evidence. Depending on the composition of the evidence set, the generated questions are categorized into the three reasoning types: \textbf{SPSR}, \textbf{SPMR}, and \textbf{MPMR}, covering single-region, single-page multi-region, and multi-page multi-region reasoning.

\textbf{Automatic Answer Verification.}
To reduce noise introduced during automatic data generation, we perform an additional answer verification stage. Given the selected EUs and the generated question, two other vision-language models independently generate answers without access to the original answer. A QA instance is retained only if the verified answers are consistent with the original answer and the question can be answered solely from the selected EUs.

\textbf{Human Quality Audit.}
To assess the quality of the automatically constructed training set, we randomly sample 500 QA instances together with their associated evidence regions for manual review. Each instance is evaluated on four criteria: evidence-region validity, question clarity, answer correctness, and evidence sufficiency. Two annotators independently review each instance, with disagreements resolved by a third annotator. Inter-annotator agreement is measured using Cohen's $\kappa$.

The automatically generated samples achieve pass rates of 96.8\%, 97.6\%, 96.2\%, and 95.4\% on the four criteria, respectively, with an overall acceptance rate of 91.8\%. Compared with the manually annotated benchmark (93.4\%), the difference is only 1.6 percentage points (95\% CI: $[-4.8,\,1.6]$), and the inter-annotator agreement reaches $\kappa=0.86$. 

These results indicate that the automatically generated training data approaches the quality of manual annotations and provides reliable supervision for large-scale model training.

\begin{table*}[t]
\centering
\caption{
Main results on the SciEGQA benchmark.
Task 1 evaluates visual grounding via the Grounding–Crop–then–Answer pipeline,
while Task 2 evaluates QA accuracy under different evidence granularities.
Bold and underline denote the best and second-best results; $\uparrow$ indicates accuracy gains over the previous granularity.
More details are available on our project webpage.
}
\label{tab:main_results}

\setlength{\tabcolsep}{5.5pt}
\renewcommand{\arraystretch}{1.08}

\newcommand{\gain}[1]{\textsuperscript{\textcolor{blue}{\scriptsize$\uparrow$#1}}}
\resizebox{\linewidth}{!}{
\begin{tabular}{l|cccccc|ccc}
\toprule
& \multicolumn{6}{c|}{\textbf{Task 1: Grounding-Crop-then-Answer}}
& \multicolumn{3}{c}{\textbf{Task 2: Evidence-Granularity-QA}} \\
\cmidrule(lr){2-7} \cmidrule(lr){8-10}
Model & Mean IoU & IoU@0.3 & IoU@0.5 & IoU@0.7 & Valid Ratio & Acc. & Document & Page & Region \\
\midrule

Qwen3-VL-8B
& 21.43\% & 24.77\% & 12.14\% & 7.27\% & 87.06\% & 32.04\%
& 9.30\% & 60.51\%\gain{51.20} & 65.87\%\gain{5.36} \\

Qwen3-VL-32B
& 31.53\% & 44.55\% & 18.55\% & 11.95\% & 98.15\% & 46.46\%
& 16.14\% & 73.14\%\gain{56.99} & 75.54\%\gain{2.40} \\

Qwen3-VL-235B-A22B
& 25.22\% & 33.52\% & 10.72\% & 4.87\% & 88.66\% & 38.63\%
& 40.30\% & 70.49\%\gain{30.19} & 75.42\%\gain{4.93} \\

Qwen3.5-27B
& 36.04\% & 47.50\% & \underline{29.39}\% & \underline{19.90}\% & 99.82\% & \underline{51.26}\%
& 42.95\% & 69.69\%\gain{26.74} & 77.02\%\gain{7.33} \\

Qwen3.5-122B-A10B
& 30.76\% & 38.08\% & 21.87\% & 14.11\% & 92.48\% & 46.95\%
& 37.58\% & 68.95\%\gain{31.36} & 73.01\%\gain{4.07} \\

Qwen3.5-397B-A17B
& \textbf{38.89}\% & 52.19\% & \textbf{32.96}\% & \textbf{20.64}\% & 99.51\% & 50.15\%
& 36.23\% & 68.08\%\gain{31.85} & 73.75\%\gain{5.67} \\

InternVL3-38B
& 5.23\% & 3.39\% & 0.62\% & 0.12\% & 89.03\% & 5.05\%
& 10.23\% & 48.18\%\gain{37.95} & 64.14\%\gain{15.96} \\

GLM-4.6V-Flash
& 19.90\% & 21.26\% & 9.67\% & 5.85\% & 77.94\% & 35.92\%
& 50.16\% & 72.03\%\gain{21.87} & 77.94\%\gain{5.91} \\

\midrule

Ernie-5
& 14.57\% & 19.10\% & 9.55\% & 4.00\% & 84.90\% & 22.12\%
& 38.51\% & 77.39\%\gain{38.88} & 81.89\%\gain{4.50} \\

Kimi-K2.5
& 7.17\% & 7.16\% & 5.49\% & 2.23\% & 96.30\% & 7.46\%
& \textbf{68.02}\% & \underline{79.98}\%\gain{11.95} & 81.21\%\gain{1.23} \\

Claude-4.6-Sonnet
& 15.93\% & 17.25\% & 4.81\% & 0.68\% & 93.84\% & 19.29\%
& 43.99\% & 68.45\%\gain{24.46} & 74.31\%\gain{5.85} \\

GPT-5.2
& 31.63\% & 49.11\% & 18.55\% & 4.81\% & \underline{99.88}\% & 41.22\%
& \underline{65.43}\% & \textbf{81.39}\%\gain{15.96} & \textbf{85.15}\%\gain{3.76} \\

\midrule
\rowcolor{yellow!20}
Qwen3-VL-8B-SFT
& \underline{38.66}\% & \textbf{57.92}\% & 28.77\% & 17.25\% & \textbf{99.94}\% & 49.97\%
& 32.90\% & 69.43\%\gain{36.56} & 76.89\%\gain{7.46} \\

\rowcolor{yellow!20}
GLM-4.6V-Flash-SFT
& 37.23\% & \underline{54.16}\% & 27.05\% & 14.23\% & \underline{99.88}\% & \textbf{52.80}\%
& 63.09\% & 78.37\%\gain{15.28} & \underline{85.10}\%\gain{6.73} \\

\bottomrule
\end{tabular}
}
\end{table*}

\subsection{SciEGQA Statistics}

The main statistics of the SciEGQA dataset are summarized in Table~\ref{tab:dataset_stats}. 
SciEGQA consists of a manually annotated \textit{fine-grained benchmark} and a large-scale automatically constructed \textit{training set}. 
The benchmark contains \textbf{1,623 question-answer pairs} with precise evidence region annotations, while the training set is generated through an automated data construction pipeline and includes approximately \textbf{30K question-answer pairs} to support large-scale model training.

Figure~\ref{fig:data-analysis} presents the statistical distributions of the dataset across multiple dimensions. 
The results show that the evidence annotations in SciEGQA cover diverse visual elements, including text blocks, figures, and tables, and exhibit a diverse spatial distribution across the page layout.

Overall, these statistics indicate that solving the SciEGQA task requires models not only to perform document-level question answering, but also to identify semantically relevant evidence regions within complex scientific documents and conduct reasoning across multiple regions or even multiple pages. 
This makes SciEGQA the first scientific document QA dataset that simultaneously supports \textit{semantic region grounding} and \textit{multi-page reasoning}.

\section{Experiment Setup}
\subsection{Evaluation Metrics}
We employ \textbf{Answer Accuracy} as the primary evaluation metric. Additionally, \textbf{Intersection over Union (IoU)} is utilized to assess the precision of predicted bounding boxes. 

\textbf{Evidence grounding.}
Valid Ratio is the proportion of predicted page-box pairs that can be parsed with a valid page index and coordinates satisfying $0\leq x_1<x_2\leq1000$ and $0\leq y_1<y_2\leq1000$.
For each sample, predicted and ground-truth boxes are grouped by page. On page $p$, we compute their pairwise IoUs and use dynamic programming to find the one-to-one maximum-IoU matching
\begin{equation}
M_p^*=\arg\max_{M_p}\sum_{(i,j)\in M_p}
\operatorname{IoU}(P_{p,i},G_{p,j}).
\label{eq:max-iou-matching}
\end{equation}
Each box can be matched at most once. We normalize by
\[
d_x=\max\left(\sum_p|P_p|,\sum_p|G_p|,1\right).
\]
The sample score is
\begin{equation}
\operatorname{MeanIoU}(x)
=\frac{1}{d_x}\sum_p\sum_{(i,j)\in M_p^*}
\operatorname{IoU}(P_{p,i},G_{p,j}),
\label{eq:mean-iou}
\end{equation}
\begin{equation}
\operatorname{IoU@}\tau(x)
=\mathbb{I}[\operatorname{MeanIoU}(x)\geq\tau],
\label{eq:iou-at-tau}
\end{equation}
where $\tau\in\{0.3,0.5,0.7\}$. Unmatched predictions and references contribute zero through $d_x$, penalizing both redundant and missing boxes. Mean IoU is averaged over samples with successfully parsed predictions; IoU@$\tau$ is averaged over all samples, with parse failures counted as zero.

\textbf{Answer accuracy.}
Exact string matching is inadequate because many questions require interpretation. Five distinct LLM judges independently compare the question, reference answer, and model answer, each casting a binary vote $v_k\in\{0,1\}$.

An answer is correct when at least three judges vote positively:
\begin{equation}
\operatorname{Acc}(x)=\mathbb{I}\!\left[\sum_{k=1}^{5}v_k\geq3\right].
\label{eq:answer-accuracy}
\end{equation}
The judges accept semantically equivalent paraphrases and numerical forms, but reject contradictory or incomplete answers. Answer Accuracy is the mean of $\operatorname{Acc}(x)$ over all samples.

\subsection{Task Definitions}

We design \textbf{two evaluation tasks} on the SciEGQA benchmark to systematically evaluate a diverse set of Vision-Language Models (VLMs), using \textbf{zero-shot inference} for base models and separately evaluating task-specific SFT checkpoints.

\textbf{Task 1: Grounding-Crop-then-Answer.}
Given the ground-truth evidence page(s) and the corresponding question, the model is required to predict a set of \textbf{bounding boxes} that localize the regions relevant to answering the question. This oracle-page setting isolates region localization from page retrieval. Each bounding box is represented as $(x_1, y_1, x_2, y_2)$, with coordinates normalized to the page coordinate space of $[0,1000]$.

Based on the predicted bounding boxes, we crop the corresponding regions from the page images and feed the cropped regions into a fixed question answering model, \textbf{Qwen3.5-27B}, to generate the final answer. Since all methods share the same answering model, this pipeline controls for differences in reasoning capability across models. We evaluate the grounding quality of the predicted regions using both \textbf{IoU} and the resulting \textbf{answer accuracy}.

\textbf{Task 2: Evidence-Granularity-QA.}
We first evaluate model performance under different levels of input granularity. 
Specifically, we provide the model with three types of inputs:
(1) the \textbf{entire document},
(2) the \textbf{evidence page(s)} containing the answer, and
(3) the \textbf{cropped evidence region(s)}.
By comparing answer accuracy across these different input settings, we analyze how different evidence granularities influence the question answering capability of VLMs.

\subsection{Evaluation Suite}
We evaluate a comprehensive range of state-of-the-art VLMs, including both open-source and closed-source models. This includes the \textbf{Qwen3-VL}~\cite{bai2025qwen3vltechnicalreport} series (8B, 32B, and the 235B-A22B variant) and the \textbf{Qwen3.5}~\cite{qwen2026qwen35} series (27B, 122B-A10B, and 397B-A17B). Furthermore, we evaluate \textbf{InternVL3-38B}~\cite{zhu2025internvl3}, \textbf{Ernie-5}~\cite{wang2026ernie}, \textbf{Kimi-K2.5}~\cite{team2026kimi}, \textbf{Claude-4.6-Sonnet}~\cite{anthropic2026sonnet46} and \textbf{GPT-5.2}~\cite{openai2025gpt52}.

To evaluate the effectiveness of the training dataset, we perform supervised fine-tuning on \textbf{Qwen3-VL-8B} for two tasks separately using the \textbf{LLaMA-Factory} framework~\cite{zheng2024llamafactory}, and additionally fine-tune \textbf{GLM-4.6V-Flash}~\cite{zai2025glm46vflash} on the same SciEGQA training set. Their task-specific checkpoints are denoted as \textbf{Qwen3-VL-8B-SFT} and \textbf{GLM-4.6V-Flash-SFT}, respectively. Task 1 retains the fixed Qwen3.5-27B answerer; Task 2 evaluates each model's own QA performance.


Models with fewer than 38B parameters run locally, while larger models are accessed via official APIs. For fair comparison, all models are evaluated using greedy decoding without sampling, and each question is answered with a single inference pass.

\begin{figure}
    \centering
    \includegraphics[width=\linewidth]{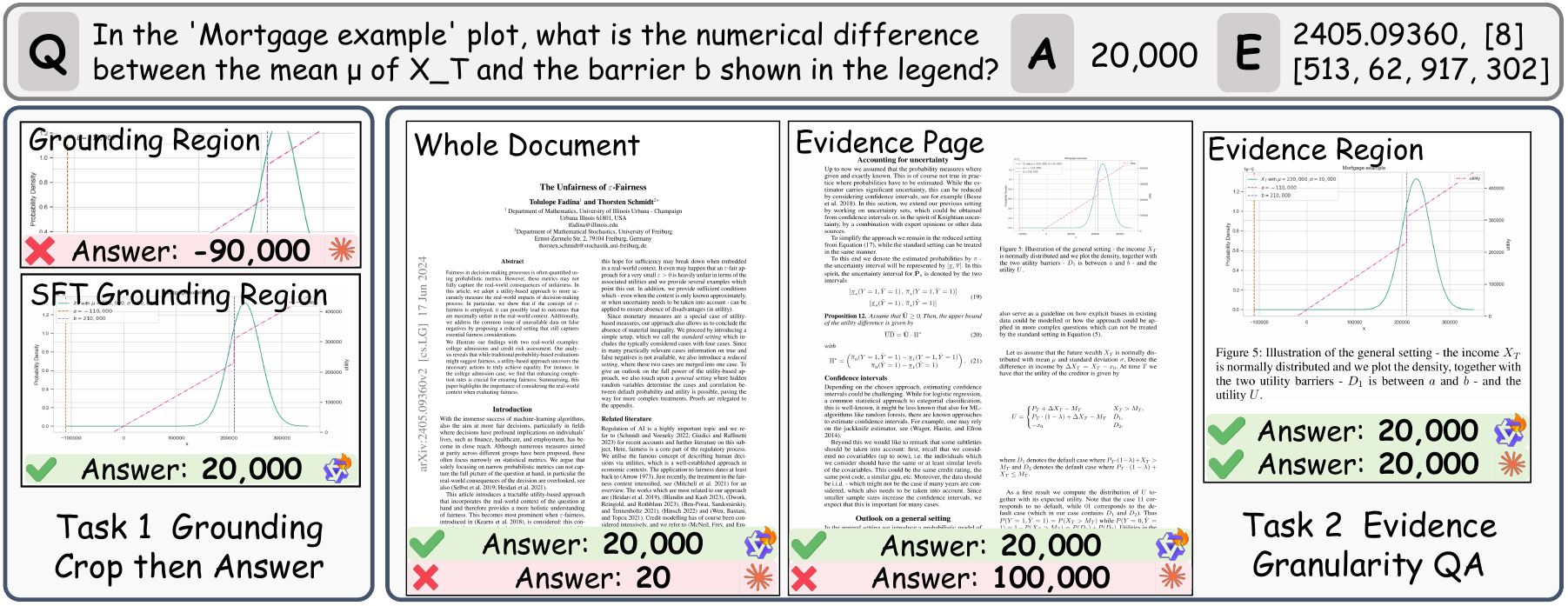}
    \caption{An SPSR case study of Claude-4.6-Sonnet and Qwen3-VL-8B-SFT on \textbf{Task 1} Grounding–Crop–then–Answer and \textbf{Task 2} Evidence-Granularity-QA.}
    \Description{A mortgage-plot question has a reference answer of 20,000. In Task 1, the Claude crop leads to an incorrect answer, while the Qwen fine-tuned crop leads to 20,000. In Task 2, Claude answers incorrectly with the full document and evidence page, but both models answer correctly with the evidence crop.}
    \label{fig:case_study}
\end{figure}

\section{Results and Analysis}
Table~\ref{tab:main_results} presents the performance of various VLMs on the SciEGQA benchmark. 
Overall, current state-of-the-art models still struggle to simultaneously achieve reliable evidence localization and accurate question answering in scientific documents, indicating that SciEGQA remains a challenging benchmark.

\textbf{Grounding Capability.}
The left part of Table~\ref{tab:main_results} evaluates visual grounding and reasoning using the \textbf{grounding-crop-then-answer} pipeline. 
Most models achieve relatively low IoU scores, with the best model reaching only \textbf{38.89\%} Mean IoU, indicating that precise evidence localization in scientific documents remains challenging. 
Since answering scientific questions often requires reasoning over specific document regions, inaccurate grounding can easily disrupt the reasoning process. 
Moreover, QA accuracy generally improves with better IoU, suggesting that inaccurate evidence localization is a key factor limiting model performance.

\textbf{Error Decomposition.}
Our controlled evaluation separates errors caused by evidence localization from those caused by downstream reasoning.
When ground-truth evidence regions are provided, the fixed answering model Qwen3.5-27B achieves 77.02\% accuracy.
In contrast, the best Grounding-Crop-then-Answer result reaches only 52.80\%, leaving a 24.22-point gap associated with imperfect localization and cropping.
Meanwhile, the remaining 22.98\% error under ground-truth regions indicates that precise evidence alone does not eliminate failures in scientific interpretation and reasoning.
Thus, current models are limited by both evidence localization and evidence-conditioned reasoning, rather than by either component alone.

\textbf{Effect of Evidence Granularity.}
The right part of Table~\ref{tab:main_results} reports QA accuracy under three evidence granularities. 
Across these three input settings, \textbf{more focused evidence leads to higher QA accuracy}.
When the entire document is provided as input, all models achieve relatively low accuracy, indicating that locating relevant evidence within large-scale context remains challenging for current models.
These results highlight the importance of precise evidence localization in scientific document QA, as it enables models to focus on relevant context and perform more reliable reasoning. 
Further analysis shows that multi-page, multi-region questions generally remain more challenging.
Appendix~\ref{app:fine-grained-comparison} additionally compares SR inputs with OCR text and layout-block images.

\textbf{Effectiveness of the Training Data.}
After fine-tuning on the SciEGQA Training Set, Qwen3-VL-8B-SFT achieves \textbf{38.66\%} Mean IoU and \textbf{49.97\%} QA accuracy under the grounding-crop-then-answer evaluation. 
Meanwhile, under \textbf{Evidence-Granularity-QA}, the model also shows consistent improvements across all evidence granularities compared with the base model. 
GLM-4.6V-Flash also improves after SFT: Mean IoU rises from 19.90\% to 37.23\%, and Task 1 QA accuracy from 35.92\% to 52.80\%, with gains under all three Task 2 input granularities. These results support the effectiveness of SciEGQA supervision beyond the Qwen family.

\textbf{Case Study.}
Figure~\ref{fig:case_study} illustrates how evidence granularity and grounding accuracy affect QA performance. 
In \textbf{Task 1}, inaccurate grounding leads to incorrect cropped regions and consequently wrong answers. 
After fine-tuning on SciEGQA, Qwen3-VL-8B-SFT correctly localizes the evidence region and produces the correct answer, while Claude-4.6-Sonnet fails.  
In \textbf{Task 2}, Claude-4.6-Sonnet fails under document- and page-level inputs due to distracting context, but succeeds when the precise evidence region is provided. 
This example highlights both the challenging nature of SciEGQA and the effectiveness of the training data.

\section{Conclusion}

In this paper, we introduce \textbf{SciEGQA}, a dataset for visual evidence-grounded question answering on scientific documents, which provides \textbf{semantic region-level evidence annotations} in the form of bounding boxes. The dataset consists of a human-annotated benchmark and a large-scale automatically constructed training set, and supports multiple reasoning scenarios in scientific documents.
Experimental results show that current Vision-Language Models still struggle with accurate evidence localization and evidence-grounded reasoning in scientific documents. We hope SciEGQA can serve as a challenging benchmark to promote future research on evidence-grounded multimodal reasoning and scientific understanding in complex documents. Meanwhile, the SciEGQA training set can serve as a valuable resource for improving the reasoning capabilities of Vision-Language Models, including future reinforcement learning–based optimization.

\section*{Ethical Considerations}
SciEGQA is constructed from publicly accessible scientific papers and does not intentionally collect private or sensitive data. To respect copyright, source documents are redistributed only when permitted by their licenses; otherwise, we release derived annotations and source identifiers.
SciEGQA is intended for research on scientific document understanding and evidence grounding. Its outputs should not be treated as verified scientific claims or used for high-stakes decisions without expert review.

\bibliographystyle{ACM-Reference-Format}
\bibliography{ref}

\appendix
\section{Benchmark Construction Details}
\label{app:benchmark-construction}

\subsection{Document Selection and Split Integrity}
\label{app:document-selection}

SciEGQA contains 80 scientific papers and 1,941 rendered pages. We sample 10 papers from each of eight domains: computer science, economics, electrical engineering and systems science, mathematics, physics, quantitative biology, quantitative finance, and statistics.

The collection comprises 17 journal articles, 12 conference papers, and 51 preprints, with both single- and double-column layouts. Before annotation, each paper is checked for stable page rendering and sufficient scientific content for evidence-grounded QA.

Each released record retains the document identifier, domain, question, reference answer, evidence-page indices, absolute and normalized bounding boxes, and evidence-modality labels. Page images are organized by domain and document identifier to preserve provenance.

We include publicly accessible papers from the target domains that contain verifiable scientific claims, experiments, derivations, quantitative analyses, or structured figures and tables. Papers are excluded if rendering fails, key content is missing or illegible, or the material is non-research or largely duplicates another candidate.

We also perform version-level deduplication. Potential arXiv, conference, and journal versions are matched using titles, authors, abstracts, and DOIs, followed by manual comparison of the main text and section structure. For the same study, we retain only the public version with the most complete results and stable layout.

All splits are isolated at the document level rather than the question, page, or evidence-region level. Test papers do not overlap with documents used for training, prompt construction, or development, and different versions of the same work are never assigned to different splits.

Finally, we check for contamination using title-author matching and text-similarity retrieval. Test QA pairs are manually compared with public QA datasets. Evidence-page and region identifiers are also checked for uniqueness and traceability.

\subsection{Semantic Region Annotation Guidelines}
\label{app:annotation-guidelines}

The benchmark is constructed without automatically generated questions or answers. Annotators first identify Semantic Regions (SRs) without reference to any question. Domain experts then write questions and associate the required SRs with each answer. This ordering prevents region boundaries from being tailored to a particular question.

For each page, annotators identify a semantic anchor, such as a paragraph, list, figure, table, or equation, and apply three operational checks. The \emph{completeness check} expands a candidate when a caption, title, note, legend, or explanatory sentence is necessary for interpretation. The \emph{minimality check} removes adjacent material that can be excluded without changing the meaning of the unit. The \emph{separation check} prevents unrelated columns or neighboring components from being merged.

Text SRs generally correspond to complete paragraphs or lists. Figure SRs include the figure and the caption or legend required to interpret it. Table SRs include the title, body, and necessary notes, while equation SRs include the surrounding text needed to identify the variables and the role of the equation. Cross-page evidence is represented by multiple page-local SRs, each preserving a complete semantic unit; one bounding box is never allowed to cross pages.

Table~\ref{tab:sr-boundary-example} gives a representative boundary decision. It illustrates that geometric tightness alone is insufficient: a plot-only box may be visually compact but semantically incomplete when its caption defines the experimental setting.

\begin{table}[t]
\centering
\small
\caption{Example of semantic-region boundary decisions.}
\label{tab:sr-boundary-example}
\begin{tabular}{@{}p{0.38\columnwidth}p{0.15\columnwidth}p{0.37\columnwidth}@{}}
\toprule
\textbf{Candidate} & \textbf{Action} & \textbf{Reason} \\
\midrule
Plot only & Expand & The caption is required to identify the setting. \\
Plot and caption & Retain & The unit is complete and minimal. \\
Plot, caption, and neighboring paragraph & Shrink & The paragraph discusses an unrelated experiment. \\
\bottomrule
\end{tabular}
\end{table}

\subsection{QA Authoring and Evidence Linking}
\label{app:qa-authoring}

After the SR pool is finalized, a domain expert selects a scientific fact or relationship to query, writes a concise reference answer, and links every SR required to derive that answer. An SR selected for a particular question becomes an Evidence Unit (EU). The expert verifies that the question has one unambiguous answer, every selected EU is necessary, no required evidence is omitted, and the answer does not depend on unselected content or external knowledge.

Questions supported by one EU are labeled Single-Page Single-Region (SPSR). Questions requiring multiple EUs on the same page are labeled Single-Page Multi-Region (SPMR), while questions requiring EUs from different pages are labeled Multi-Page Multi-Region (MPMR). Thus, the reasoning type is determined by the finalized evidence set rather than by surface cues in the question. The benchmark contains 46.15\% SPSR, 34.26\% SPMR, and 19.59\% MPMR instances.

The released annotation uses the following core fields: \texttt{query} and \texttt{answer} store the QA pair; \texttt{doc\_name} and \texttt{category} identify the source paper; \texttt{evidence\_page} records the relevant pages; \texttt{rel\_bbox} stores normalized coordinates; and \texttt{subimg\_type} records whether an EU is text, a table, or an image. Bounding boxes are grouped first by evidence page and then by region, preserving the structure required for SPMR and MPMR instances.

\subsection{Annotation Agreement and Adjudication}
\label{app:annotation-agreement}

Each benchmark paper is independently processed by two annotators. Corresponding SR proposals are matched and their Intersection over Union (IoU) is computed. Cases with IoU below 0.9 are sent to a third annotator for adjudication. The adjudicator determines whether the disagreement is caused by missing context, excessive surrounding content, or an incorrect merge or split, and the adjudicated boundary replaces the two initial proposals in the final SR pool.

Question-answer instances are also checked for answer correctness and evidence sufficiency. In particular, the reviewer confirms that the selected EUs are jointly sufficient and that the reference answer can be derived without consulting any unselected region.


\section{Automatic Training-Set Construction Details}
\label{app:automatic-construction}

Implementation details of the automated construction pipeline are available in our \href{https://github.com/yuwenhan07/SciEGQA}{GitHub repository}; here, we clarify several key aspects of training-set construction.

\subsection{Pipeline and Model Configuration}
\label{app:pipeline-configuration}

The automatic construction pipeline processes each paper in five stages: (1) PDF rasterization, (2) candidate-region segmentation, (3) semantic-region screening and deduplication, (4) question-answer generation and independent verification, and (5) conversion to the released JSONL format. All PDF pages are rendered as PNG images at 300 dpi before visual processing.

Following the procedure described in the main text, we use SAM3 to generate candidate masks with the default checkpoint configuration and inference parameters provided by its official implementation, and subsequently convert these masks into rectangular proposals. Before semantic screening, proposals whose bounding-box area covers less than 5\% or more than 70\% of the page are removed.

For each sample, the VLM responsible for semantic-region screening and question-answer generation is randomly selected from a predefined pool consisting of \texttt{Qwen3-VL-235B-A22B}, \texttt{GPT-5}, and \texttt{Gemini-3}. The selected model checks whether each crop forms a complete and minimal semantic unit and assigns it to one of three modality categories: \texttt{text}, \texttt{table}, or \texttt{image}. During independent verification, the other two VLMs independently answer the generated question using only the selected EUs, without access to the generated answer. We retain a sample only when both verification answers are consistent with the generated answer and the evidence is sufficient, as described in the main text.

\subsection{Prompt Templates}
\label{app:construction-prompts}

These templates cover region screening and region-aware QA generation. Bracketed expressions specify the evidence scope; image tokens and serialization details are omitted.

\begin{listingbox}[label={box:crop-screening-prompt}, title={Crop-screening template}, colback=lightgreen]
Determine whether the supplied crop is a complete and sufficiently fine-grained semantic region. Set \texttt{keep=true} only if it contains one complete logical unit, includes the context required for interpretation, and excludes unrelated content. Reject blank, illegible, truncated, or mixed-content crops. Classify the primary modality as \texttt{text}, \texttt{table}, or \texttt{image}. Return only \texttt{\{"keep": true/false, "type": "..."\}}; if uncertain, set \texttt{keep=false}.
\end{listingbox}

\begin{listingbox}[label={box:region-aware-qa-prompt}, title={Region-aware QA-generation template}]
You are given one or more evidence regions extracted from a scientific document. Region boundaries are represented by one of the following scopes:

\texttt{[SPSR: page $p$; regions $r_1$]}

\texttt{[SPMR: page $p$; regions $r_1,\ldots,r_n$]}

\texttt{[MPMR: page $p_1$; regions $r_1,\ldots,r_i$;}\\
\texttt{\hspace*{1em}$\ldots$; page $p_m$; regions $r_j,\ldots,r_n$]}

Use only the content inside the specified scope. Generate three diverse question--answer pairs that match its structure. For SPSR, one region must be sufficient. For SPMR, the answer must require multiple regions on the same page. For MPMR, it must require evidence from multiple regions across multiple pages. Do not generate a multi-region question if any single region is sufficient.

Prefer questions involving comparisons, numerical trends, relationships, hierarchy, or explicit scientific statements. For each pair, return \texttt{question}, a concise \texttt{answer}, \texttt{region\_scope}, an answer type from \texttt{[span, number, date, yesno, categorical]}, and a difficulty label from \texttt{[easy, medium, hard]}. If the evidence does not support a valid pair, return \texttt{NO\_VALID\_QA}.
\end{listingbox}

\subsection{Filtering and Failure Handling}
\label{app:failure-handling}

The construction code applies conservative filtering at each stage. Candidate crops that fail the area constraint are removed before any VLM call. Crop-screening output is parsed as strict JSON; malformed or unrecognized outputs default to \texttt{keep=false} and are assigned the \texttt{unknown} modality rather than being retained.

After semantic screening, highly overlapping crops of the same known modality are deduplicated. For two regions whose intersection covers at least 90\% of the smaller region, the smaller crop is retained for text, reflecting minimality, while the larger crop is retained for tables and figures, reflecting completeness. Overlapping regions with different or unknown modality labels are preserved to avoid deleting potentially valid evidence.

If a batch fails during screening or generation, the pipeline processes its images individually where possible. Unreadable images, missing crop files, invalid boxes, and unusable evidence crops are skipped and recorded. The parser accepts tagged, JSON-like, and line-based QA variants, removes duplicate questions, and discards entries without both a question and an answer. Raw generation outputs are retained alongside parsed records for later inspection.


\section{Evaluation and Implementation Details}
\label{app:evaluation-details}

\subsection{Input Formatting and Inference Prompts}
\label{app:inference-prompts}

The two main benchmark tasks operate on page images, not OCR transcripts or parsed text. Each instance uses a ShareGPT-style multimodal format containing image tokens, an instruction, and a question. The code builds four views from each annotation: \texttt{document} supplies every page in document order; \texttt{page} supplies the annotated evidence pages; \texttt{crop} supplies the ground-truth evidence regions; and \texttt{bbox} supplies the evidence pages and asks the model to predict normalized region coordinates.

\begin{listingbox}[label={box:qa-inference-prompt}, title={QA prompt}, colback=black!5]
Answer the question using only the document image(s). Return only the final answer with no explanation.
\end{listingbox}

\begin{listingbox}[label={box:grounding-inference-prompt}, title={Grounding prompt}, colback=black!5]
Locate the region or regions needed to answer the question in each document image. Return only JSON grouped by page in the format $[[[x_1,y_1,x_2,y_2],\ldots],\ldots]$. Each outer item corresponds to one input image or page in order, and every coordinate is normalized to $[0,1000]$.
\end{listingbox}

For Task~1, predicted regions are cropped from the supplied evidence pages and passed, together with the original question, to the fixed Qwen3.5-27B answering model. For Task~2, the same QA instruction is used for document-, page-, and region-level inputs so that the primary controlled variable is evidence granularity.

\subsection{Grounding Output Parsing and Region Cropping}
\label{app:grounding-parser}

The evaluator first attempts to parse grounding output as JSON and also supports JSON enclosed in code fences, Python-style list literals, coordinate strings, and dictionaries with common keys such as \texttt{bbox} or \texttt{coordinates}. Page-grouped predictions are matched only against references on the corresponding page. When a flat prediction contains exactly one box for each page of a multi-page single-box example, it is promoted to page structure in input order; otherwise, the evaluator applies global flat matching and records this fallback mode. An output that cannot be parsed is counted as a parse failure and contributes zero to IoU-threshold rates; Mean IoU is averaged over successfully parsed samples.

Within each page, predicted and reference boxes are matched one-to-one to maximize the sum of IoUs. The sample denominator is the larger of the number of predicted and reference boxes, so both missing and redundant regions reduce the score. IoU@0.3, IoU@0.5, and IoU@0.7 report the fractions of all benchmark samples whose normalized sample-level scores reach the corresponding thresholds.

For the crop-then-answer stage, normalized coordinates are mapped back to source-image dimensions and clamped to valid boundaries. The released reconstruction utility rejects crops whose width or height is smaller than 28 pixels and does not generate answers for samples with no usable crop. Such samples remain in the Task 1 QA denominator of 1,623 and are counted as incorrect. Multiple valid crops are retained as separate images and are presented in page order and then region order.

\subsection{LLM-as-a-Judge Protocol}
\label{app:llm-judge}

Strict string matching cannot reliably evaluate answers that allow semantically equivalent expressions, alternative numerical forms, or different wording. We therefore adopt an LLM-as-a-Judge protocol for automatic answer evaluation.

Each judge receives the complete question, reference answer, and candidate response, and independently returns a binary decision. The instruction used by the released evaluation utility is shown below.

\begin{listingbox}[label={box:llm-judge-prompt}, title={LLM-as-a-Judge prompt}, colback=black!5, breakable=false]
Evaluate whether the model response correctly answers the question based on the reference answer. The output must be strictly either \texttt{correct} or \texttt{incorrect}. Accept semantically equivalent answers and reject responses that contradict the reference, omit required information, introduce a substantive error, or fail to answer the question. Return only one word.
\end{listingbox}

For the final evaluation, we use five distinct LLMs as independent judges to reduce bias from any single evaluator: \texttt{DeepSeek-V4}, \texttt{Gemini-3}, \texttt{GPT-5}, \texttt{Claude-4.6}, \texttt{Qwen3-235B-A22B}.

All judges receive identical inputs and use the same evaluation instruction and generation configuration, ensuring comparability across evaluators.

Let $v_k(x)\in\{0,1\}$ denote the decision of judge $k$ for sample $x$, where 1 indicates correct and 0 indicates incorrect. Sample-level correctness is determined by majority vote:
\begin{equation}
\operatorname{Acc}(x)=\mathbb{I}\!\left[\sum_{k=1}^{5}v_k(x)\geq3\right].
\label{eq:appendix-judge-vote}
\end{equation}
Because five judges are used, ties cannot occur. Dataset-level Answer Accuracy is the mean of these binary sample-level labels.

\paragraph{Human Validation of Answer Scoring.}
On 200 randomly sampled test instances, human judgments agreed with this majority vote on 184 instances (92.0\%, Cohen's $\kappa=0.802$). Human annotators independently assessed each candidate response.

\subsection{Supervised Fine-Tuning Details}
\label{app:sft-details}

We use LLaMA-Factory to fine-tune Qwen3-VL-8B and GLM-4.6V-Flash on the same SciEGQA training set. For each model, we train two separate checkpoints: one for grounding and one for QA. These task-specific checkpoints are denoted as Qwen3-VL-8B-SFT and GLM-4.6V-Flash-SFT, respectively, in the main paper. Both model families use LoRA-based parameter-efficient fine-tuning with the same training settings. Training is performed with the AdamW optimizer and a cosine learning-rate scheduler, using an initial learning rate of $1\times10^{-3}$. Each checkpoint is trained for 3 epochs with an effective batch size of 64 on 8 NVIDIA A100 GPUs with 80GB memory each.

Training examples follow the same multimodal message format and task instructions used during evaluation. Optimization settings and data formatting are kept consistent across both model families and the grounding and QA tasks. For Task~1, each grounding checkpoint predicts regions that are passed to the fixed Qwen3.5-27B answerer; Task~2 evaluates the corresponding QA checkpoint under document-, page-, and region-level inputs.


\section{Additional Results and Analysis}
\label{app:additional-results}

This section provides complementary analyses of the two benchmark tasks. We examine how localization quality propagates to answer accuracy, how the amount of visual context affects QA, and how these effects vary across the three reasoning structures in SciEGQA.

\subsection{Grounding Quality and Answer Accuracy}
\label{app:grounding-answer-analysis}

\begin{figure*}[t]
    \centering
    \includegraphics[width=\textwidth]{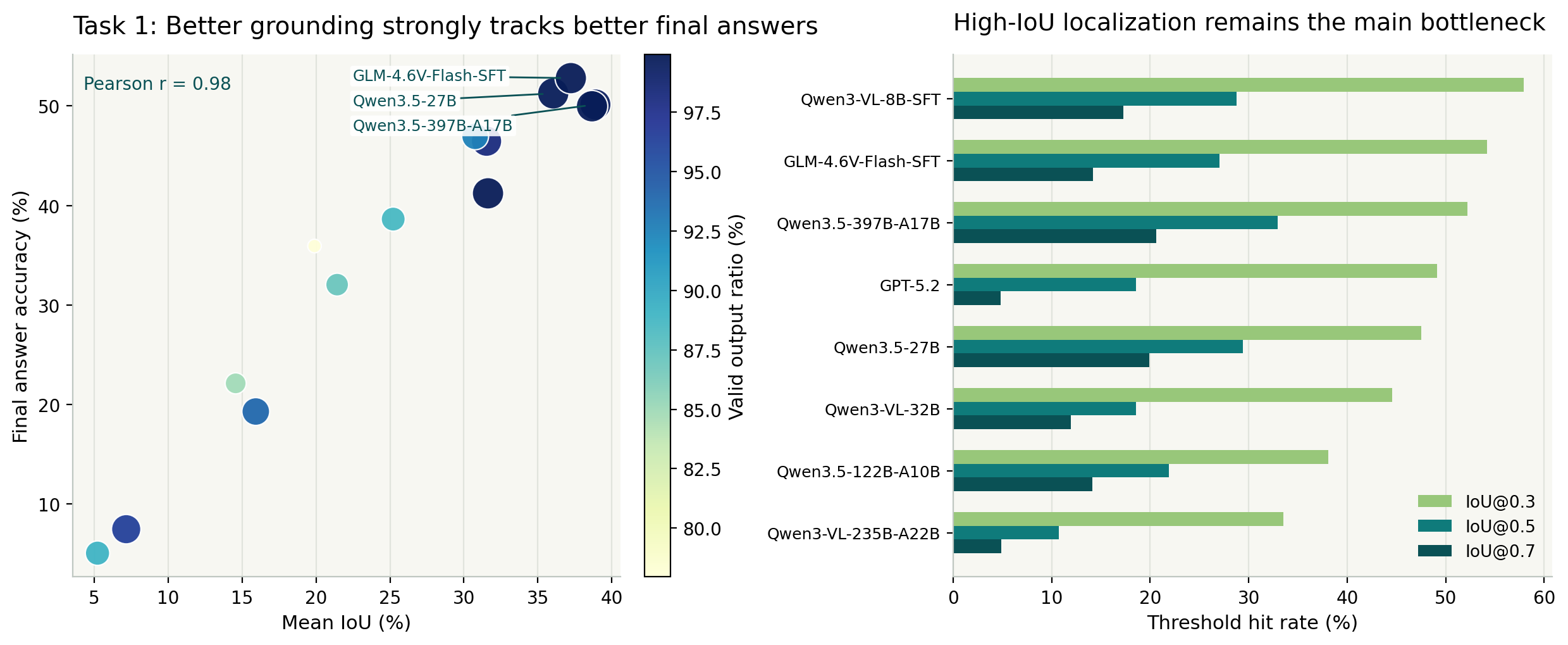}
    \caption{Task~1 grounding analysis. Left: mean IoU is strongly correlated with downstream answer accuracy across evaluated models; marker color indicates the valid-output ratio. Right: localization hit rates at IoU thresholds of 0.3, 0.5, and 0.7 for the eight models with the highest IoU@0.3.}
    \Description{A scatter plot relates mean IoU to final answer accuracy, and a grouped horizontal bar chart compares localization hit rates at three IoU thresholds.}
    \label{fig:task1-grounding-analysis}
\end{figure*}

As shown in Figure~\ref{fig:task1-grounding-analysis}, mean IoU and Task~1 answer accuracy have a Pearson correlation of $r=0.98$ across the 14 models in the main results table. The predicted boxes directly determine the visual evidence passed to the fixed QA model.

Because the downstream answering model and its prompt are fixed for all grounding systems, this comparison isolates the practical effect of the predicted crops. Improvements in end-to-end accuracy therefore depend not only on whether a model emits valid coordinates, but also on whether those coordinates preserve the evidence needed to answer the question.

When a predicted region deviates from the reference evidence, its crop may omit necessary content or introduce irrelevant information. The resulting evidence corruption can therefore turn an otherwise answerable question into an incorrect prediction.

The sharp decline from IoU@0.3 to IoU@0.7 further shows that precise localization remains substantially harder than coarse localization. Even when current models identify the general area, they fail to accurately recover semantically complete region boundaries.

The valid-output ratio and IoU capture two distinct failure modes. A syntactically valid box may still exclude a caption, legend, table note, or explanatory sentence, while a malformed output prevents a usable crop from being constructed at all. The threshold results consequently distinguish broad evidence discovery from boundary-level semantic completeness.

\subsection{Effect of Evidence Granularity}
\label{app:results-granularity}

\begin{figure}[H]
    \centering
    \includegraphics[width=\linewidth]{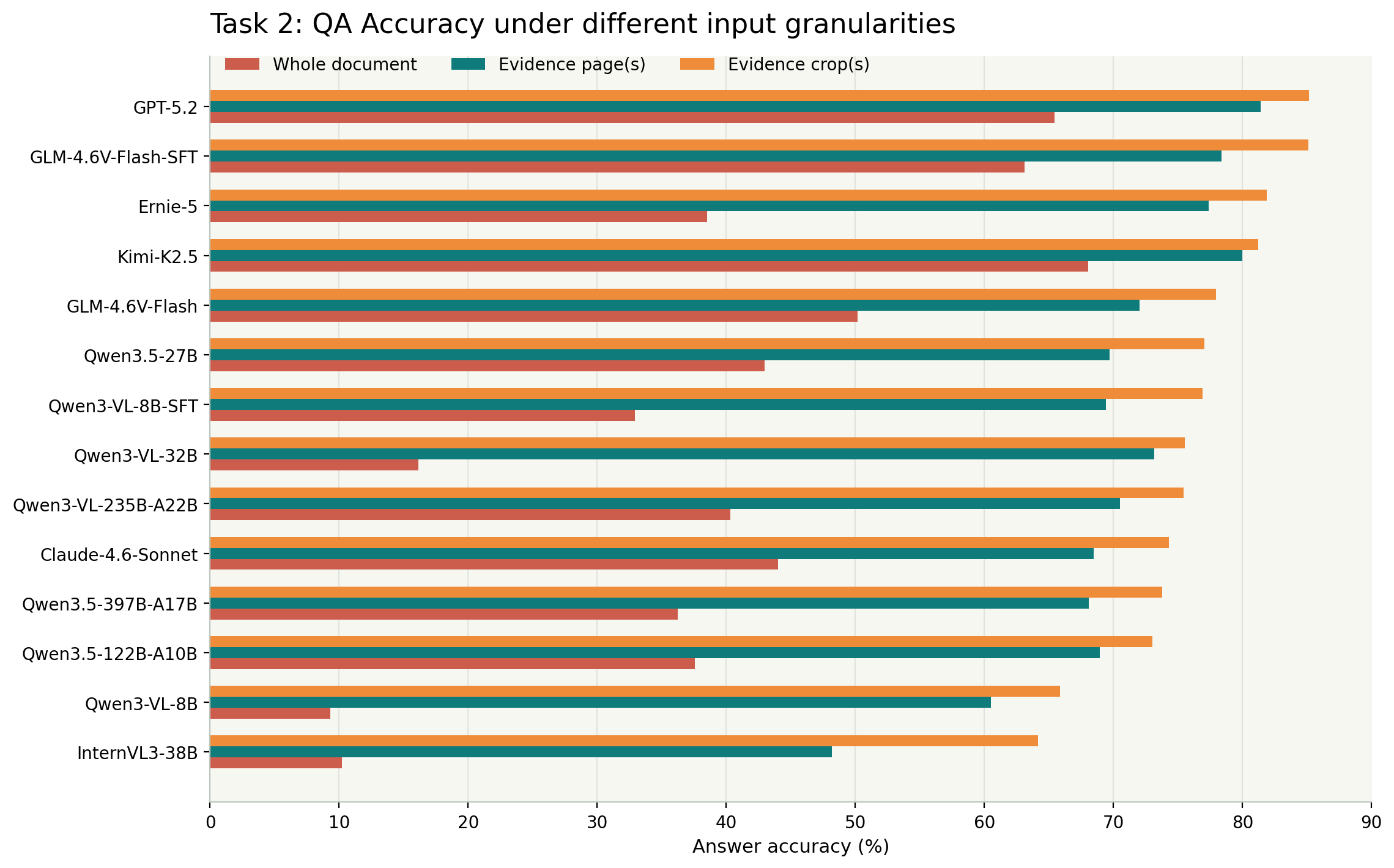}
    \caption{Task~2 answer accuracy under whole-document, evidence-page, and evidence-crop inputs. More focused evidence consistently improves QA accuracy across the evaluated models.}
    \Description{A grouped horizontal bar chart compares answer accuracy for fourteen models under whole-document, evidence-page, and evidence-crop inputs.}
    \label{fig:task2-granularity}
\end{figure}

\begin{figure*}[t]
    \centering
    \includegraphics[width=0.82\textwidth]{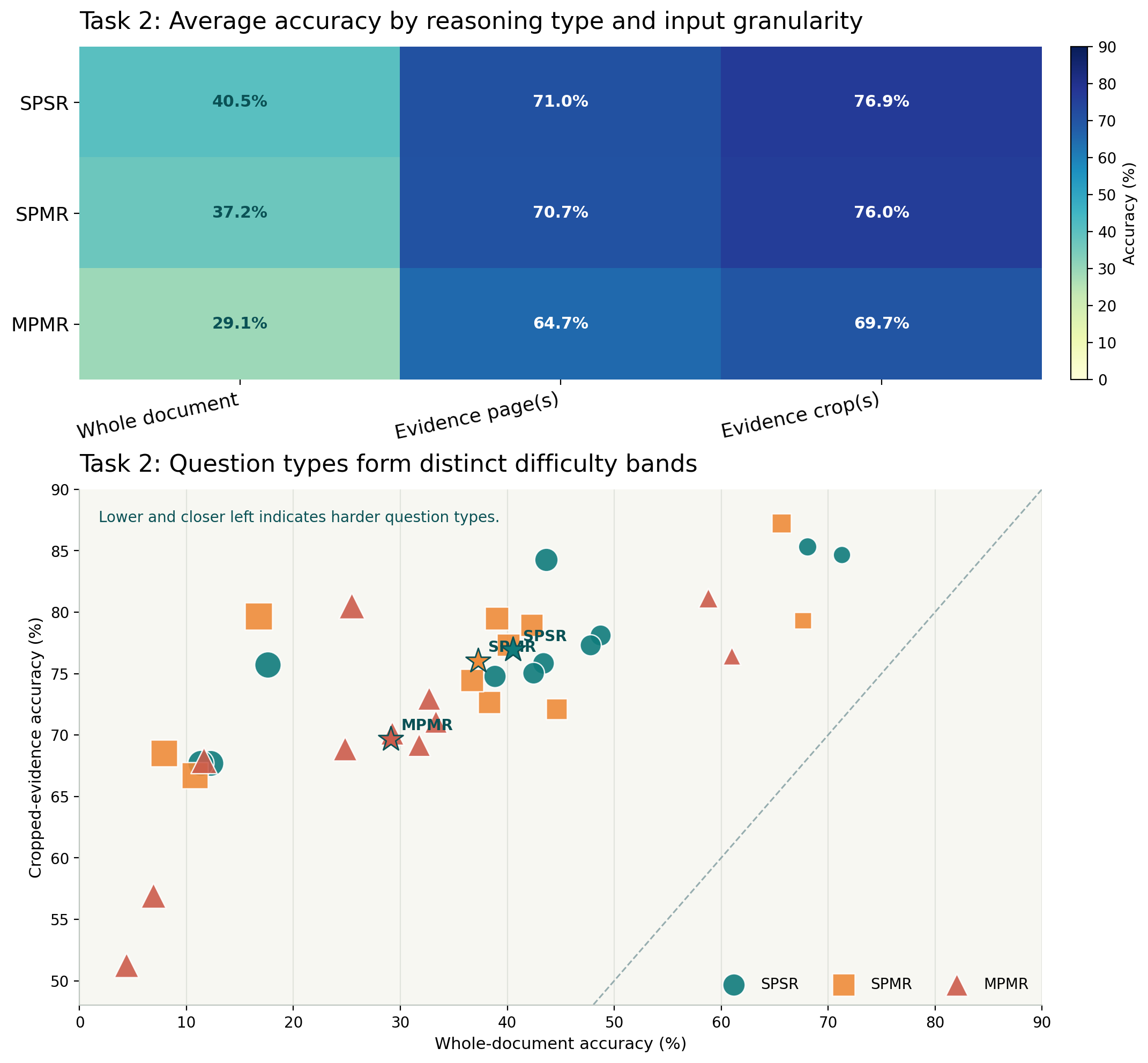}
    \caption{Task~2 results by reasoning type for the 11 models with available per-type scores; GLM and SFT models are excluded. Top: average accuracy for SPSR, SPMR, and MPMR under three input granularities. Bottom: per-model whole-document and cropped-evidence accuracy, with stars marking the average of each reasoning type.}
    \Description{A heat map and scatter plot compare SPSR, SPMR, and MPMR accuracy under whole-document, evidence-page, and evidence-crop inputs.}
    \label{fig:task2-reasoning-breakdown}
\end{figure*}

As shown in Figure~\ref{fig:task2-granularity}, QA accuracy increases as the input narrows from the whole document to the evidence page or pages and finally to the evidence crop or crops. This pattern is consistent across the evaluated models.

The three settings form a controlled continuum of evidence access. Whole-document input requires the model to search the complete paper before reasoning; evidence-page input removes cross-page retrieval but retains unrelated material on the selected pages; evidence-crop input provides the annotated semantic regions directly.

Reducing irrelevant context allows a model to attend more directly to the information needed for reasoning. The result also explains why grounding quality is consequential: accurate crops provide the strongest QA input, whereas inaccurate crops can remove that advantage.

The monotonic trend across different model families indicates that sensitivity to irrelevant visual context is not confined to a particular architecture or model scale. At the same time, accuracy remains below perfect even with ground-truth crops, showing that precise evidence access does not eliminate errors in scientific interpretation, comparison, and numerical reasoning.

\subsection{Results by Reasoning Type}
\label{app:results-reasoning-type}

Figure~\ref{fig:task2-reasoning-breakdown} shows distinct difficulty bands. Mean accuracy with evidence crops decreases from 76.9\% for SPSR to 76.0\% for SPMR and 69.7\% for MPMR as evidence aggregation becomes more complex.

The same ordering holds for whole-document input (40.5\%, 37.2\%, and 29.1\%) and evidence-page input (71.0\%, 70.7\%, and 64.7\%), showing that MPMR is especially sensitive to irrelevant context.

SPSR and SPMR become close with oracle pages or crops, suggesting that within-page integration is manageable after localization. MPMR remains lower because evidence must be preserved and combined across page boundaries.

Thus, the three categories separately test localized, within-page, and cross-page evidence integration and should not be treated as a homogeneous QA set.

Together, these analyses identify three bottlenecks: valid evidence localization, distraction control, and cross-region integration. SciEGQA therefore requires joint progress in grounding and reasoning.

\subsection{Comparison with Other Fine-Grained Evidence Representations}
\label{app:fine-grained-comparison}

We randomly sample 100 questions from each of the SPSR (single-page single-region), SPMR (single-page multi-region), and MPMR (multi-page multi-region) categories, yielding 300 questions in total. We compare six input conditions on the same questions using Qwen3.5-27B as the fixed QA model. All conditions use the annotated evidence pages as their source and share the same QA prompt, generation settings, and answer-scoring protocol.

OCR text is extracted using PaddleOCR~\cite{cui2025paddleocr}, and layout blocks are parsed using PP-StructureV3~\cite{cui2025pp}. For the layout-block images, we compare two conditions: providing all parsed blocks and providing only blocks that overlap with the annotated evidence-region bounding boxes. The gold-overlap condition retains blocks satisfying $|B\cap G|/\min(|B|,|G|)\geq0.5$ for a ground-truth evidence box $G$ and parsed block $B$; crop boundaries remain those of the parser. A single DeepSeek-V4-Flash judge scores all 300 questions in each condition, counting failed predictions as incorrect.

As shown in Table~\ref{tab:fine-grained-comparison}, SRs achieve higher QA accuracy as an evidence representation in this comparison. This result supports the value of region representations that balance localization precision and semantic completeness, and further illustrates the contribution of region-level evidence localization to multimodal QA. Differences also reflect OCR information loss, layout parsing quality, and input-budget failures.

\begin{table}[t]
\centering
\caption{QA accuracy (\%) by evidence representation on 300 questions (100 per reasoning type).}
\label{tab:fine-grained-comparison}
\setlength{\tabcolsep}{3pt}
\begin{tabular}{@{}lrrrr@{}}
\toprule
\textbf{Input condition} & \textbf{SPSR} & \textbf{SPMR} & \textbf{MPMR} & \textbf{Overall} \\
\midrule
Evidence-page images & 86 & 83 & 85 & 84.67 \\
Gold SR images & \textbf{92} & \textbf{89} & \textbf{90} & \textbf{90.33} \\
Page OCR text & 66 & 55 & 42 & 54.33 \\
Region OCR text & 71 & 60 & 44 & 58.33 \\
All layout-block images & 79 & 78 & 57 & 71.33 \\
Gold-overlap layout blocks & 89 & 84 & 85 & 86.00 \\
\bottomrule
\end{tabular}
\end{table}

\end{document}